\documentclass[graybox,oneside]{svmult}

\newcommand{\comment}[1]{} \usepackage{natbib}
\usepackage[normalem]{ulem} \usepackage{amsmath}

\usepackage{indentfirst}
\usepackage{fancyheadings}
\usepackage{listings}
\usepackage{courier}
\usepackage{color}
\usepackage{graphicx}
\usepackage{longtable}
\usepackage{amsmath}
\usepackage{url}
\usepackage{algorithm}
\usepackage{algorithmic}
\usepackage{multirow}
\usepackage[top=.8in,bottom=1in,left=1in,right=1in]{geometry}

\newcommand{\mf}[1]{{\mbox{\emph{#1}}}}
\newcommand{\smf}[1]{{\mbox{\emph{\small #1}}}}

\newcommand{\hide}[1]{} 

\bibpunct{(}{)}{,}{s}{,}{,}

\makeindex 

\begin{document}

\bibliographystyle{plainnat}

\title*{Simultaneous Optimization of Both Node and Edge Conservation in
Network Alignment via WAVE}
\author{
\textbf{\large{Yihan Sun$^{1,2,3}$, 
Joseph Crawford$^1$, Jie Tang$^2$ and 
Tijana Milenkovi\'{c}$^{1,}$*}}\\ 
$^1$Department of Computer Science and Engineering,
Interdisciplinary Center for Network Science and Applications, and ECK Institute for Global Health, University of Notre Dame. \\
$^2$Department of Computer Science and Technology, Tsinghua University.\\
$^3$Computer Science Department, Carnegie Mellon University. \\
*Corresponding Author (E-mail: tmilenko@nd.edu)
}
%
%

\maketitle

\vspace{-1.1cm}

\noindent\textbf{Abstract.} Network alignment can be used to transfer 
functional knowledge between conserved regions of different networks.
Typically, existing methods use a node cost function (NCF) to compute
similarity between nodes in different networks and an alignment
strategy (AS) to find high-scoring alignments with respect to the
total NCF over all aligned nodes (or node conservation). But, they
then evaluate quality of their alignments via some other measure that
is different than the node conservation measure used to guide the
alignment construction process. Typically, one measures the amount of
conserved edges, but only after alignments are produced. Hence, a
recent attempt aimed to directly maximize the amount of conserved
edges while constructing alignments, which improved alignment
accuracy. Here, we aim to directly maximize both node and edge
conservation during alignment construction to further improve
alignment accuracy. For this, we design a novel measure of edge
conservation that (unlike existing measures that treat each conserved
edge the same) weighs each conserved edge so that edges with highly
NCF-similar end nodes are favored. As a result, we introduce a novel
AS, \textbf{W}eighted \textbf{A}lignment \textbf{V}ot\textbf{E}r
(WAVE), which can optimize any measures of node and edge conservation,
and which can be used with any NCF or combination of multiple NCFs.
Using WAVE on top of established state-of-the-art NCFs leads to
superior alignments compared to the existing methods that optimize
only node conservation or only edge conservation or that treat each
conserved edge the same. And while we evaluate WAVE in the
computational biology domain, it is easily applicable in any domain.

\vspace{-0.6cm}

\section{Introduction}

\vspace{-0.3cm}

\subsection{Motivation}

\vspace{-0.3cm}

Network alignment aims to find topologically or functionally similar
regions between different networks.
It has applications in different areas, including computational
biology
\cite{GRAAL,HGRAAL,IsoRankN,MIGRAAL,GHOST,MilenkovicACMBCB2013,MAGNA,Faisal2014a},
ontology matching \cite{rimom,paris,sigma,flooding}, pattern
recognition \cite{pm1,pm2}, language processing \cite{symword}, social
networks \cite{bilink,crosslinking}, and others
\cite{cvcg1,cvcg2,cvcg3,cvcg4,chemical}. Our study focuses mainly on
the computational biology domain.

Protein-protein interaction (PPI) networks have been the main focus of
network alignment research among all biological networks. PPI network
alignment can be used to transfer biological knowledge from the
network of a poorly studied species to the network of a well studied
species. This is of importance because not all cellular processes can
easily be studied via biological experiments.
For example, studying aging in human has to rely on across-species
transfer of aging-related knowledge from model species
\cite{Faisal2014}. And network alignment can be (and has been) used
for this \cite{MilenkovicACMBCB2013,MAGNA,Faisal2014a}. However, the
problem is computationally intractable, as the underlying subgraph
isomorphism problem is NP-complete \cite{cook1971}. Thus, network
alignment methods are heuristics.

Network alignment can be local or global. Local network alignment aims
to align well local network regions
\cite{PathBlast,Sharan2005,Flannick2006,Mawish,Berg04,Liang2006a,Berg2006,Mina20014,AlignNemo}.
As such, it often fails to find large conserved regions between
networks. Hence, majority of recent research has focused on global
network alignment
\cite{Singh2007,Flannick2008,Singh2008,GraphM,IsoRankN,GRAAL,HGRAAL,MIGRAAL,GHOST,NETAL,Narayanan2011,Guo2009,NATALIE,NATALIE2,IsoRankN,MilenkovicACMBCB2013,Faisal2014a,MAGNA,NewSurvey2014,Crawford2014},
which can find large conserved regions between networks. Typically,
global network alignment aims to generate one-to-one node mapping
between two networks \cite{NewSurvey2014} (although exceptions exist
that produce many-to-many node mappings or that align more than two
networks \cite{IsoRankN}, but such methods are out of the scope of our
study).

Of one-to-one global network alignment methods, many consist of two
algorithmic components, namely, a node cost function (NCF) and an
alignment strategy (AS)
\cite{MilenkovicACMBCB2013,Faisal2014a,Crawford2014}. NCF captures
pairwise similarities between nodes in different networks, and AS then
searches for good alignments based on the NCF information. It has
already been recognized that when two methods of this two-component
NCF-AS type are compared, to fairly evaluate the methods, one should
mix and match the different methods' NCFs and ASs, because NCF of one
method and AS of another method could lead to a new method that is
actually superior to the original methods
\cite{MilenkovicACMBCB2013,Faisal2014a,Crawford2014}.

We base our work on established state-of-the-art NCFs of existing
methods. Then, we propose a novel AS, \textbf{W}eighted
\textbf{A}lignment \textbf{V}ot\textbf{E}r (WAVE), which when used on
top of the established NCFs leads to a new superior method for global
network alignment. And while we evaluate our new method in the
computational biology domain, the method is easily applicable in any
domain.

\vspace{-0.7cm}

\subsection{Related Work}

\vspace{-0.3cm}

We focus on NCFs of two popular existing methods, MI-GRAAL
\cite{MIGRAAL} and GHOST \cite{GHOST}, and we aim to improve with our
new WAVE AS upon these methods' ASs.

MI-GRAAL improves upon its predecessors, GRAAL \cite{GRAAL} and
H-GRAAL \cite{HGRAAL}, by using the same NCF but by combining their
ASs (see below). MI-GRAAL's NCF computes topological similarity
between extended network neighborhoods of two nodes
\cite{Milenkovic2008,Solava2012,Memisevic10b,MMGP_Roy_Soc_09}. It does
so by relying on the concept of small induced subgraphs called
graphlets (e.g., a triangle or a square)
\cite{Przulj06ECCB,GraphCrunch}, which are used to summarize the
topology of up to 4-deep network neighborhood of a node into its
graphlet degree vector (GDV)
\cite{Milenkovic2008,Hulovatyy2014,Milenkovic2011}. Then,
GDV-similarity is used as MI-GRAAL's NCF, which compares nodes' GDVs
to compute their topological similarity. MI-GRAAL also allows for
integration of other node similarity measures into its NCF, such as
protein sequence similarity. We recently showed
\cite{MilenkovicACMBCB2013,Faisal2014a} that MI-GRAAL's NCF is
superior to another, Google PageRank algorithm-based NCF, which is
used by IsoRank \cite{Singh2007} and IsoRankN \cite{IsoRankN}.
Regarding AS \cite{Crawford2014}, MI-GRAAL's AS combines GRAAL's
greedy seed-and-extend AS with H-GRAAL's optimal AS that uses the
Hungarian algorithm to solve linear assignment problem of maximizing
total NCF over all aligned nodes.

GHOST's NCF is conceptually similar to MI-GRAAL's, as it also assumes
two nodes from different networks to be similar if their neighborhoods
are similar. However, the mathematical and implementation details of
the two NCFs are different. Namely, GHOST's NCF takes into account a
node's $k$-hop neighborhood, (in this study, $k=4$). Then, its NCF
computes topological distance (or equivalently, similarity) between
two nodes by comparing the nodes' ``spectral signatures''. We
recently fairly compared MI-GRAAL's GDV-similarity-based NCF with
GHOST's ``spectral signature''-based NCF within our above
mix-and-match framework, concluding that MI-GRAAL's NCF is superior or
comparable to GHOST's NCF, depending on data \cite{Crawford2014}.
Hence, since none of the two NCFs was dominant in all cases, we
consider both NCFs in our study. Just as MI-GRAAL, GHOST also allows
for integration of protein sequence information into its NCF.
Regarding AS, GHOST is also a seed-and-extend algorithm, like
MI-GRAAL. However, GHOST's AS considers quadratic (instead of linear)
assignment problem. When we evaluated the two ASs, their performance
was data-dependent \cite{Crawford2014}. Hence, we consider both ASs
in our study.

There exist additional more recent network alignment methods
\cite{NewSurvey2014}, both those that also belong to the category of
NCF-AS methods, such as NETAL \cite{NETAL}, and those that do not,
such as MAGNA \cite{MAGNA}. These methods became available close to
completion of our study, and as such, we were not able to include them
into the design of our new method. (Hence, NETAL implements a
different NCF compared to NCFs of MI-GRAAL and GHOST, along with a
different AS compared to ASs of MI-GRAAL, GHOST, and WAVE.) However,
we still consider these methods in our evaluation. Importantly, our
goal is to show that when we use under an existing NCF (such as
MI-GRAAL's or GHOST's) our new WAVE AS, we get alignments of higher
quality compared to when using an existing AS (such as MI-GRAAL's or
GHOST's) on the same NCF. This would be sufficient to illustrate the
superiority of WAVE. If in the process we also improve upon the more
recent methods, such as those that use a different NCF and especially
those that do not belong to the NCF-AS category, that would further
demonstrate WAVE's superiority.

\vspace{-0.7cm}

\subsection{Our Contributions and Significance}\label{contributions}

\vspace{-0.3cm}

We introduce WAVE, a novel, general, and as we will show superior AS,
which can be combined with any NCF. WAVE is applicable to any domain.
We evaluate it on biological networks. 

Its novelty and significance is as follows. The existing ASs use NCF
scores to rapidly identify from possible alignments the high-scoring
alignments with respect to the overall NCF (henceforth also referred
to as node conservation). But, their alignment accuracy is then
evaluated with some other measure that is different than NCF used to
construct the alignments \cite{MAGNA}. Typically, one measures the
amount of conserved (i.e., aligned) edges. Hence, a recent attempt
aimed to directly maximize edge conservation during alignment
construction \cite{MAGNA}. Here, we aim to optimize both node and
edge conservation while constructing an alignment, as also recognized
by a recent effort \cite{NETAL}. In the process, unlike the existing
methods that treat each conserved edge the same, we aim to favor
conserved edges with NCF-similar end nodes over those with
NCF-dissimilar end nodes. And we design WAVE with these goals in
mind.

We combine WAVE with NCF of MI-GRAAL as well as with NCF of GHOST. We
denote the resulting network aligners as M-W and G-W, respectively. We
compare M-W and G-W against the original MI-GRAAL (henceforth also
denoted by M-M) and GHOST (henceforth also denoted by G-G), which use
MI-GRAAL's NCF and AS and GHOST's NCF and AS, respectively. Further,
we compare M-W and G-W with a new method introduced recently
\cite{Crawford2014}, which is the combination of GHOST's NCF and
MI-GRAAL's AS (henceforth also denoted by G-M). This allows us to
test the performance of WAVE against the performance of MI-GRAAL's and
GHOST's ASs, under each of MI-GRAAL's and GHOST's NCF. We note that
we cannot compare M-W and G-W against the combination of MI-GRAAL's
NCF and GHOST's AS (i.e., M-G), as the current implementation of GHOST
does not allow for plugging MI-GRAAL's NCF into GHOST's AS
\cite{Crawford2014}. Finally, we compare M-W and G-W against the very
recent NETAL and MAGNA methods.

We evaluate all methods on synthetic and real-world PPI networks,
relying on established data and performance measures
\cite{GRAAL,HGRAAL,MIGRAAL,GHOST,MilenkovicACMBCB2013,MAGNA,Faisal2014a}.
We find that WAVE AS is overall superior to the existing ASs,
especially in terms of topological alignment quality. Also, WAVE
overall performs comparably to or better than NETAL and MAGNA,
especially on synthetic data. This further validates WAVE, because
NETAL implements a newer and thus possibly more efficient NCF compared
to NCFs of M-W or G-W, which might give NETAL unfrair advantage over
WAVE.

\vspace{-0.7cm}

\section{Methods}
\label{sec:method}

\vspace{-0.3cm}

\subsection{Data}

\vspace{-0.3cm}

We evaluate WAVE on two popular network sets
\cite{GRAAL,HGRAAL,MIGRAAL,GHOST,MilenkovicACMBCB2013,MAGNA,Faisal2014a}:
1) ``synthetic'' networks with known node mapping, and 2) real-world
networks with unknown node mapping.

The ``synthetic'' data consists of a high-confidence yeast PPI network
\cite{Collins07} with 1,004 nodes and 8,323 PPIs, and of five noisy
networks constructed by adding to the high-confidence network a
percentage of low-confidence PPIs from the same data set
\cite{Collins07}; we vary the percentage from $5\%$ to $25\%$ in
increments of $5\%$. We align the original high-confidence network to
each of the five noisy networks, resulting in five network pairs to be
aligned. Since we know the correct node correspondence, we can
measure to what extent an aligner correctly reconstructs the
correspondence.

The real-world set contains binary (yeast two-hybrid, Y2H) PPI
networks of four species: \emph{S. cerevisiae} (yeast/Y), with 3,321
nodes and 8,021 edges, \emph{D. melanogaster} (fly/F), with 7,111
nodes and 23,376 edges, \emph{C. elegans} (worm/W), with 2,582 nodes
and 4,322 edges, and \emph{H. sapiens} (human/H), with 6,167 nodes
and 15,940 edges. We align each pair of the networks, resulting in six
pairs. If we aimed to predict new biological knowledge, we would have
evaluated our method on additional PPIs, such as those obtained via
affinity purification followed by mass spectrometry (AP/MS). However,
since our main focus is method evaluation, of all PPIs, we focus on
binary Y2H PPIs because: 1) they have been argued to be of higher
quality than literature-curated PPIs supported by a single publication
\cite{Venkatesan2009,Hulovatyy2014}, and 2) the same Y2H networks have
already been used in many existing studies
\cite{GRAAL,HGRAAL,MIGRAAL,GHOST,MilenkovicACMBCB2013,MAGNA,Faisal2014a}.
Ultimately, what is important for a fair evaluation is that all
methods are tested on the same data, be it Y2H, AP/MS, or other PPIs
\cite{Faisal2014a}.

When we combine within NCF nodes' topological similarity scores with
their sequence similarity scores (see below), for the latter, we rely
on BLAST bit-values from the NCBI database \cite{BLASTscore}. When we
evaluate biological alignment quality with respect to functional
enrichment of the aligned nodes (see below), we rely on Gene Ontology
(GO) data \cite{GOref} to evaluate the biological alignment quality.
We use the same data versions as in our recent work
\cite{MilenkovicACMBCB2013,Faisal2014a,Crawford2014}.

\vspace{-0.7cm}

\subsection{Combining Topological and Sequence Information Within NCF}
\label{combining}

\vspace{-0.3cm}

We compute the linear combination of topological node similarity
scores $s_{t}$ and sequence node similarity scores $s_{s}$ of nodes
$u$ and $v$ as: $s(u,v) = \alpha
s_{\smf{t}}(u,v)+(1-\alpha)s_{\smf{s}}(u,v)$. We vary $\alpha$ from
from 0.0 to 1.0 in increments of 0.1. We do this for all combinations
of MI-GRAAL's, GHOST's, and WAVE's NCFs and ASs. When we compare WAVE
to recent NETAL and MAGNA, since current implementations of NETAL and
MAGNA do not support inclusion of sequence information, for these
methods, we only study topology-based alignments (corresponding to
$\alpha$ of 1).

\vspace{-0.7cm}

\subsection{Evaluation of Alignment Quality}

\vspace{-0.3cm}

If we align graph $G(V_G,E_G)$ to graph $H(V_H,E_H)$ (where
$|E_G|\le|E_H|$) via an injective function $f: V_G \rightarrow V_H$,
let us denote with $E'_G$ this edge set: $ E'_G = \{(f(u),f(v)) | u
\in V_G, v \in V_G, (u,v) \in E_G\} $. Also, let us denote with
$E'_H$ the edge set of the subgraph of $H$ that is induced on nodes
from $V_H$ that are images of nodes from $V_G$. $E'_H = \{(f(u),f(v))
| u \in V_G, v \in V_G, (f(u),f(v)) \in E_H\}$. With these notations
in mind, we next define alignment quality measures that we use.

\vspace{-0.5cm}

\subsubsection{Topological Alignment Quality Measures}
\label{evaluation}

\vspace{-0.2cm}



\noindent\emph{{Node correctness (NC)}}. Given a known true node
mapping (which is typically not available in real-world applications),
NC is the percentage of node pairs that are correctly mapped by an
alignment. If $f^*: V_G \rightarrow V_H$ is the correct node mapping
of $G$ to $H$ and $f: V_G \rightarrow V_H$ is an alignment produced by
the aligner, then $NC = \frac{|\{u \in V_G : f^*(u) = f(u) \}|}{|V_G|}
\times 100\%$ \cite{GRAAL}.

\noindent\emph{Edge Correctness (EC)}. EC represents the percentage of
edges from $G$, the smaller network (in terms of the number of
nodes), which are aligned to edges from $H$, the larger
network\,\cite{GRAAL}. Formally, $EC = \frac{|E'_{G} \cap
E'_H|}{|E_{G}|}\times{100\%}$, where the numerator is the number
of conserved edges.

\noindent\emph{{Induced conserved structure (ICS)}}. ICS is defined as
$ICS = \frac{|E'_{G} \cap E'_H|}{|E'_H|}\times{100\%}$. It was
introduced because EC fails to penalize for misaligning edges in the
larger network, i.e., $E'_H$, as EC is defined with respect to edges
in $E_G$ only \cite{GHOST}. Hence, ICS accounts for this. However,
ICS now fails to penalize for misaligning edges in the smaller
network, i.e., $E_G$, as it is defined with respect to edges in
$E'_H$ only. Hence, the following measure, S$^3$, was introduced
recently to penalize for misaligning edges in both the smaller and
the larger network \cite{MAGNA}.

\noindent\emph{{Symmetric substructure score (S$^3$)}}. S$^3$ is
defined as $S^3 = \frac{|E'_{G} \cap E'_H|}{|E_G| +|E'_H| - |E'_{G}
\cap E'_H|}\times{100\%}$ \cite{MAGNA}. Thus, S$^3$ keeps
advantages of both EC and ICS while addressing their drawbacks. And
it was already demonstrated to be the superior of the three measures
\cite{MAGNA}. Thus, we discard EC and ICS measures from further
consideration, and instead, we report results for S$^3$.

\noindent The size of the \emph{{largest connected common subgraph
(LCCS)}}\cite{GRAAL}. In addition to counting aligned edges via
S$^3$ measure, it is important that the aligned edges cluster
together to form large, dense, and connected subgraphs, rather than
being isolated. In this context, a connected common subgraph (CCS)
is defined as a connected subgraph (not necessarily induced) that
appears in both networks \cite{HGRAAL}. We measure the size of the
largest CCS (LCCS) in terms of the number of nodes as well as edges,
as defined in the MAGNA paper \cite{MAGNA}.

In summary, we focus on NC, S$^3$, and LCCS. The larger their
values, the better the topological alignment quality.

\vspace{-0.5cm}

\subsubsection{Biological Alignment Quality Measures}

\vspace{-0.2cm}

To transfer function from well annotated network regions to poorly
unannotated ones, which is the main motivation behind network
alignment in computational biology, alignment should be of good
biological quality, mapping nodes that perform similar function.

\noindent\emph{Gene Ontology Enrichment (GO).} One could measure GO,
the percentage of aligned protein pairs in which the two proteins
\emph{share} at least one GO term, out of all aligned protein pairs
in which both proteins are annotated with at least one GO term
\cite{MAGNA,Crawford2014}. In this case, complete GO annotation data
is used, independent of GO evidence code.

\noindent\emph{Experimental GO (Exp-GO).} However, since many GO
annotations have been obtained via sequence comparison, and since the
aligners use sequence information within their NCF, it is important to
test the aligners when considering only GO annotation data with
experimental evidence codes. This avoids the circular argument of
evaluating alignment quality with respect to the same data that was
used to construct the alignments
\cite{GRAAL,HGRAAL,MIGRAAL,MAGNA,Faisal2014a,Crawford2014}. Thus, we
discard GO measure from further consideration, and instead, we report
results for Exp-GO.

In summary, we focus Exp-GO. The larger its value, the better the
biological alignment quality.

\vspace{-0.6cm}

\subsection{Our Methodology}

\vspace{-0.2cm}

\subsubsection{Problem Definition}
\label{sec:problem}

\vspace{-0.2cm}

Existing network alignment methods aim to maximize either node
conservation or edge conservation. Further, they treat each conserved
edge the same. Here, we aim to simultaneously maximize both node and
edge conservation, while favoring conserved edges whose end nodes are
highly similar. Given a measure of node conservation (denoted as Node
Alignment Quality, NAQ) and a measure of (weighted) edge conservation
(denoted as Edge Alignment Quality, EAQ), our goal is to optimize the
following expression (denoted as Alignment Quality, AQ):

\vspace{-0.2cm}

\begin{equation}
\label{objective}
AQ(G,H,f) = \beta_n \mf{NAQ}(G, H, f) + \beta_e \mf{EAQ}(G, H, f),
\end{equation}

where $\beta_n$ and $\beta_e$ are parameters used to balance between
NAQ and NEQ.

As a proof of concept, we use the following measures as NAQ and EAQ
(although any other measure can be used instead). We use the sum of
NCF scores over all aligned pairs as our NAQ, which we denote as
weighted node conservation (WNC). We design a novel measure of edge
conservation as our EAQ, as follows. Similar to EC, ICS, and S$^3$,
this new measure counts the number of conserved edges, but unlike EC,
ICS, or S$^3$ that treat each conserved edge the same, our new measure
weighs each conserved edge by the NCF-based similarity of its end
nodes, so that aligning an edge with highly similar end nodes is
preferred over aligning an edge with dissimilar end nodes. We denote
our new EAQ measure as weighted edge conservation (WEC).

Formally, we define WNC and WEC as follows. Given a pairwise node
similarity matrix $s$ with respect to the given NCF, we denote
similarity between $u\in V_G$ and $v\in V_H$ in this matrix as
$s_{uv}$. Also, we represent the injection $f: V_G \rightarrow V_H$ as
a matrix $y_{|V_G|\times|V_H|}$, where $y_{ij}=1$ if and only if
$f(i)=j$ and $y_{ij}=0$ otherwise. Thus, the matrix satisfies the
following three constraints:

\vspace{-0.6cm}

\begin{equation}
\label{constrain3}
y_{ij}\in \{0,1\}, ~~ \forall i \in V_G, \forall j \in V_H; ~~~~~~
\sum_{l=1}^{|V_H|} y_{il} \le 1, ~~ \forall i \in V_G; ~~~~~~
\sum_{l=1}^{|V_G|} y_{lj} \le 1, ~~ \forall j \in V_H
\end{equation}

Then:

\vspace{-0.7cm}

\begin{equation}
\label{WNCF}
\mf{WNC}=\sum_{i\in V_G}\sum_{j\in V_H} y_{ij}s_{ij}
\end{equation}

To formally define WEC, recall the definitions of EC, ICS, and S$^3$
(Section \ref{evaluation}). All three measures have the same
numerator, which we can now rewrite as:

\vspace{-0.3cm}

\begin{equation}
\label{traditionalECy}
|E'_{G} \cap E'_H| = \frac{1}{2} \sum_{i\in V_G} \sum_{j\in V_H} \sum_{k \in \mathcal{N}_{i}} \sum_{l \in \mathcal{N}_{j}} y_{ij}y_{kl}
\end{equation}

Here, $\mathcal{N}_{i}$ denotes the neighborhood of node $i$, i.e.,
the set of nodes connected to $i$. Since each conserved edge will be
counted twice, the $1\over 2$ constant corrects for this.

Now, to leverage the weight of conserved edges by the NCF-based
similarity of its end nodes (see above), we define WEC as follows:

\vspace{-0.3cm}

\begin{equation}
\label{WEC}
\mf{WEC}=\sum_{i\in V_G} \sum_{j\in V_H} \sum_{k \in \mathcal{N}_{i}} \sum_{l \in \mathcal{N}_{j}} y_{ij}y_{kl}s_{kl}
\end{equation}

With WNC as our NAQ and WEC as our EAQ, formally, our problem is to
find a matrix $y$ that satisfies Eqn. \ref{constrain3} and maximizes
the following objective function:
\begin{align}
\nonumber
AQ(G,H,y)=&\beta_n \mf{NAQ} + \beta_e \mf{EAQ}
\nonumber
=\beta_n \mf{WNC}+\beta_e \mf{WEC}\\
\label{objection}
= &\beta_n \sum_{i\in V_G}\sum_{j\in V_H} y_{ij}s_{ij} 
+ \beta_e \sum_{i\in V_G} \sum_{j\in V_H} \sum_{k \in \mathcal{N}_{i}} \sum_{l \in \mathcal{N}_{j}} y_{ij}y_{kl}s_{kl} 
\end{align}

Optimizing the WNC part in Eqn. \ref{objection} is solvable in
polynomial time (e.g., by using Hungarian algorithm for maximum
bipartite weighted matching). However, optimizing the whole function
on general graphs is NP-hard. We propose WAVE to solve this problem,
while allowing for trade off between node conservation and edge
conservation (as the two might not always agree).

\vspace{-0.5cm}

\subsubsection{Weighted Alignment VotEr (WAVE)}

\vspace{-0.2cm}

We set $\beta_n = \beta_e = 1$, to equally favor WNC and WEC.
Evaluating other combinations of the parameters is of interest but is
not of our primary focus. Then, we can rewrite Eqn. \ref{objection}
as:

\vspace{-0.3cm}

\begin{equation}
AQ(G,H,y)=\sum_{(i,j)\in V_{G}\times V_{H}} y_{ij}\left(s_{ij}+\sum_{(k,l)\in \mathcal{N}_{i}\times \mathcal{N}_{j}}y_{kl}s_{kl} \right)
\end{equation}

Next, we use set $A=\{(u,v)\,|\,u \in V_G,v\in V_H,y_{uv}=1\}$ to
denote our alignment, so our objective function has set $A$ as a
variable. Then, we use a greedy approach to maximize the objective
function, as follows. We start with an empty alignment set $A_0$. In
each step $t$, given the current alignment $A_{t-1}$, we calculate the
marginal gain of adding an available node pair $(u,v)$ (in the sense
that so far $v$ and $u$ are both unaligned) into $A$. (For a function
$f(S)$ with variable $S$ as a set, the marginal gain of adding an
element $e$ into $S$ is defined as $f(S\cup \{e\})-f(S)$.) That is, we
calculate: $AQ(A_{t-1}\cup \{(u,v)\})-AQ(A_{t-1})$. Then, we align
the pair $(u^*,v^*)$ with the highest marginal gain, i.e., $A_t =
A_{t-1} \cup \{(u^*,v^*)\}$. To calculate the marginal gain
efficiently, we keep the current marginal gain of each node pair and
update it in each step. The marginal gain of the node pair $(u,v)$ to
$AQ$ is $s_{uv}$ at the beginning (when $A$ is empty, if we align this
pair, we can only get $s_{uv}$ in WNC part). In each step, note that
if we align two nodes $u \in V_G$ and $v \in V_H$, the side effect is
that, in the following steps, when we align another pair of nodes $u'
\in \mathcal{N}_{u}, v' \in \mathcal{N}_{v}$, both the similarity of
$(u,v)$ and $(u',v')$ will be counted once more by the correctly
linked edge, namely, the edge $(u,u')\in E_G$ and $(v,v')\in E_H$.
Thus, the marginal gain of $(u',v')$ will be $s_{uv}+s_{u'v'}$ more
after $(u,v)$ is aligned.

Intuitively, this process is like voting. When a pair of nodes is
aligned, this node pair has a chance to vote for their neighbors: when
$u$ and $v$ are aligned, all other node pairs in
$\mathcal{N}_{u}\times \mathcal{N}_{v}$ receive a weighted vote (with
weight $s_{uv}+s_{u'v'}$) from $(u,v)$, and the weight consists of two
parts: 1) the ``authority'' of the voter, i.e., $s_{uv}$, 2) the
``certainty'' of the votee, i.e., $s_{u'v'}$.

The weight for the initial votes of each node pair is the original
$s_{uv}$ (which forms the WNC part in the objective function). In
every round of WAVE, node pair $(u^*,v^*)$ with the highest vote is
aligned, and $(u^*,v^*)$ then vote for all the pairs in
$\mathcal{N}_{u^*}\times\mathcal{N}_{v^*}$. The current vote that a
node pair gets from its aligned neighbors is the marginal gain to
objective function of aligning them.

The WAVE pseudocode is shown the Appendix.

\vspace{-0.7cm}

\section{Results and Discussion}
\label{sec:result}

\vspace{-0.3cm}

We evaluate five aligners resulting from mixing and matching NCFs of
MI-GRAAL and GHOST with ASs of MI-GRAAL, GHOST, and WAVE: M-M, M-W,
G-M, G-G, and G-W (Section \ref{contributions}). Also, we evaluate
WAVE (the best of M-W and G-W) against NETAL and MAGNA.

By comparing M-M and M-W, we can directly and fairly evaluate ASs of
MI-GRAAL and WAVE under MI-GRAAL's NCF. By comparing G-M, G-G, and
G-W, we can directly and fairly evaluate ASs of MI-GRAAL, GHOST, and
WAVE under GHOST's NCF. If WAVE AS produces better alignments compared
to the existing methods' ASs under both of the existing NCFs, this
would indicate WAVE's superiority. If WAVE also produces better
alignments compared to NETAL and MAGNA, this would even further
demonstrate WAVE's superiority. However, this is not a strict
requirement, as the two new methods either implement both different
(newer, and thus possibly superior) NCF than any of M-W and G-W as
well as different AS (in case of NETAL), which might give them an
unfair advantage, or they work on different principles (in case of
MAGNA) and could be thus viewed as complementary to WAVE.

For each combination of network pair, value of $\alpha$ (denoting
topological versus sequence information within NCF), and alignment
quality measure (Section \ref{sec:method}), we do the following.

First, to extract the most out of each source of biological
information, it would be beneficial to know how much of new biological
knowledge can be uncovered solely from topology before integrating it
with other sources of biological information, such as protein sequence
information \cite{GRAAL,HGRAAL,MIGRAAL}. Thus, we first compare the
different NCF-AS methods on topology-only alignments (corresponding to
$\alpha$ of 1 within NCF). Also, since NETAL and MAGNA also produce
topology-only alignments, here, we can compare WAVE to these methods.

Second, we examine different contributions of topology versus sequence
information in NCF (by varying $\alpha$), and for each method, we
choose the best value of $\alpha$, i.e., the method's best alignment.
Since current implementations of NETAL and MAGNA do not allow for
inclusion of sequence information, here, we cannot compare WAVE to
these methods.

For ``synthetic'' (noisy yeast) networks with known node mapping, we
report alignment quality with respect to NC, S$^3$, LCCS, and Exp-GO.
For real-world PPI networks of different species with unknown node
mapping, we report alignment quality with respect to S$^3$, LCCS, and
Exp-GO.

\vspace{-0.7cm}

\subsection{Comparison of the Five NCF-AS Methods}

\vspace{-0.3cm}

Here, we compare M-M, M-W, G-M, G-G, and G-W, to test whether WAVE AS
improves upon ASs of MI-GRAAL and GHOST under the same (MI-GRAAL's or
GHOST's) NCF.

\vspace{-0.4cm}

\subsubsection{Networks With Known Node Mapping}

\vspace{-0.2cm}

\begin{figure*}[h]
\centering
\begin{tabular}{ccccc}
\hspace{-0.7cm} \includegraphics[width=0.3\columnwidth]{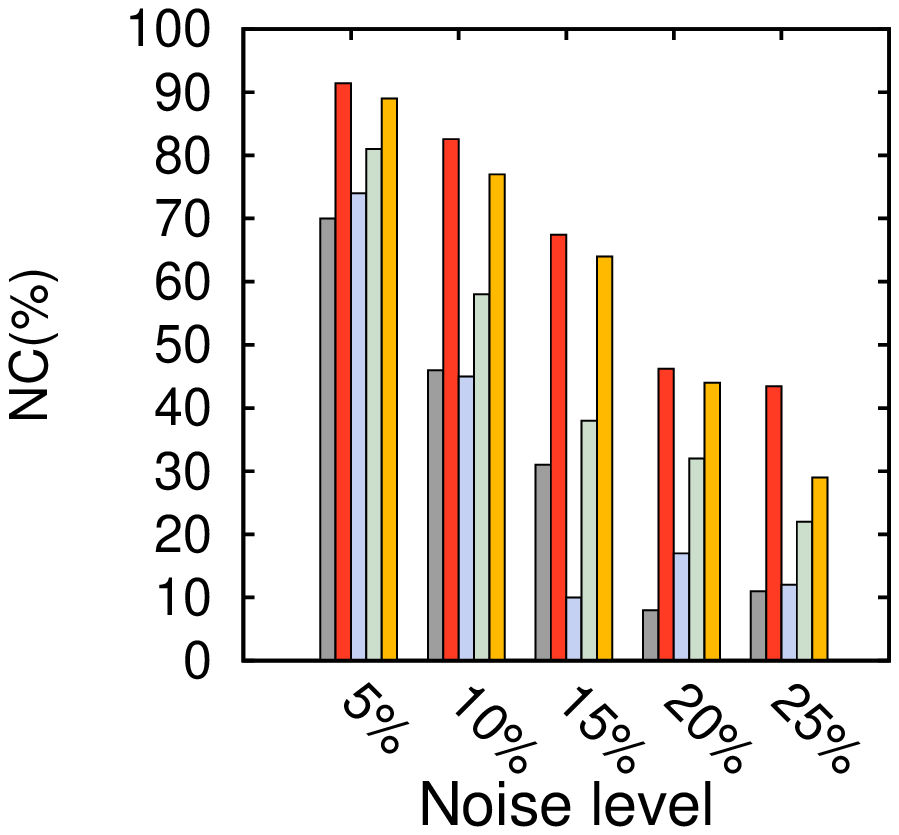}&
\hspace{-1.25cm} \includegraphics[width=0.3\columnwidth]{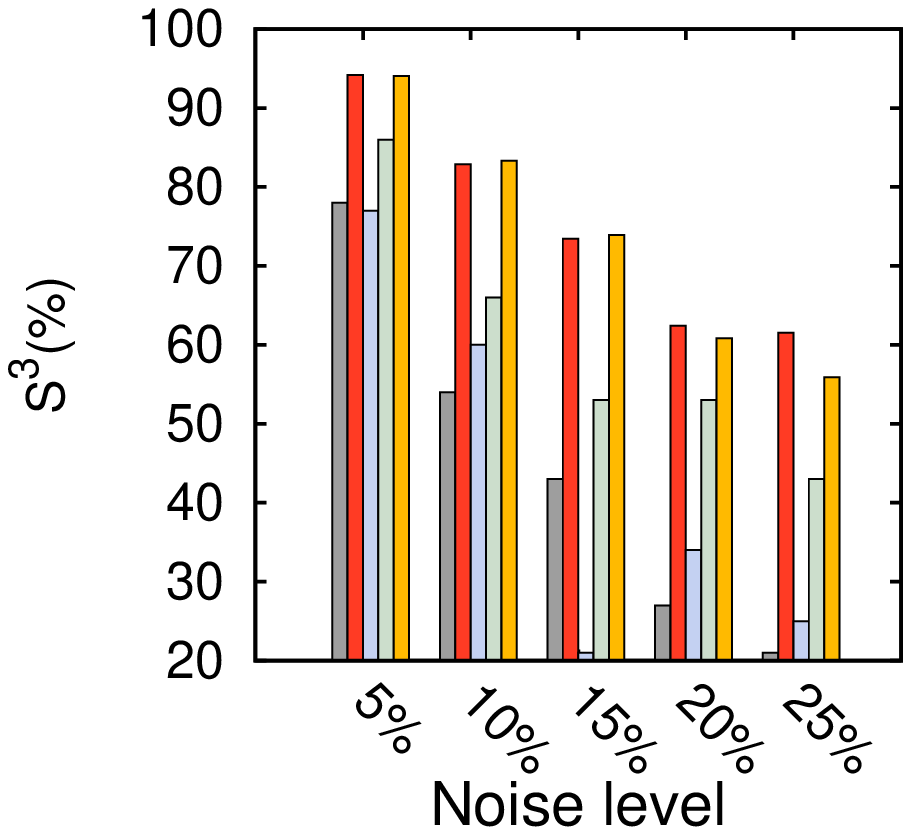}&
\hspace{-1.25cm} \includegraphics[width=0.3\columnwidth]{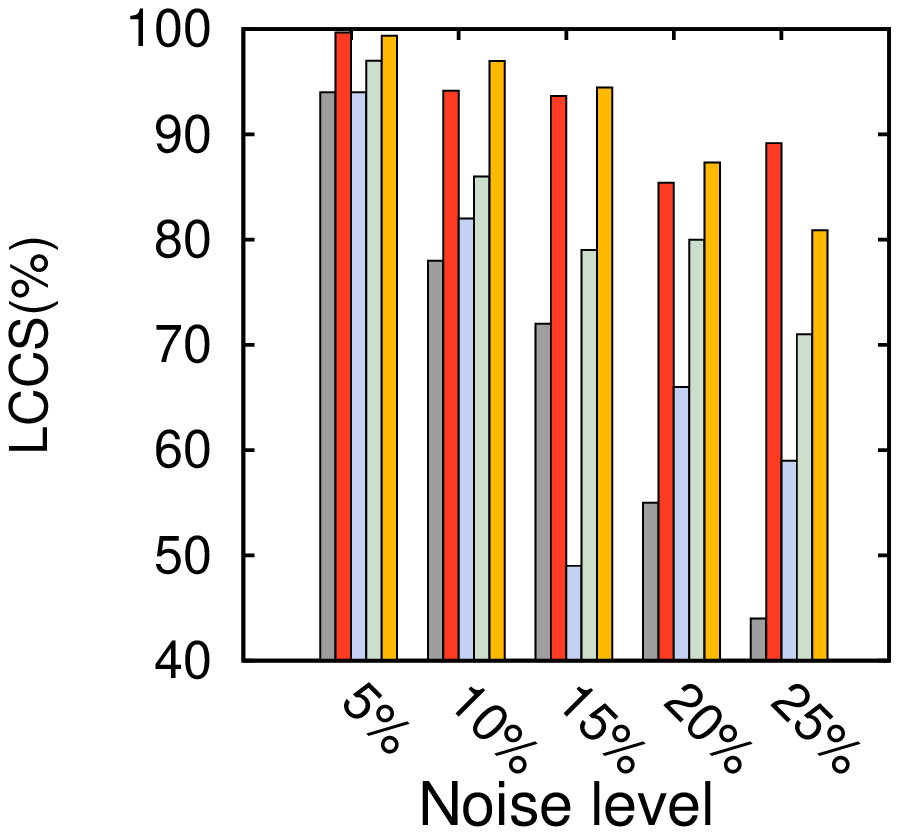}&
\hspace{-1.25cm} \includegraphics[width=0.3\columnwidth]{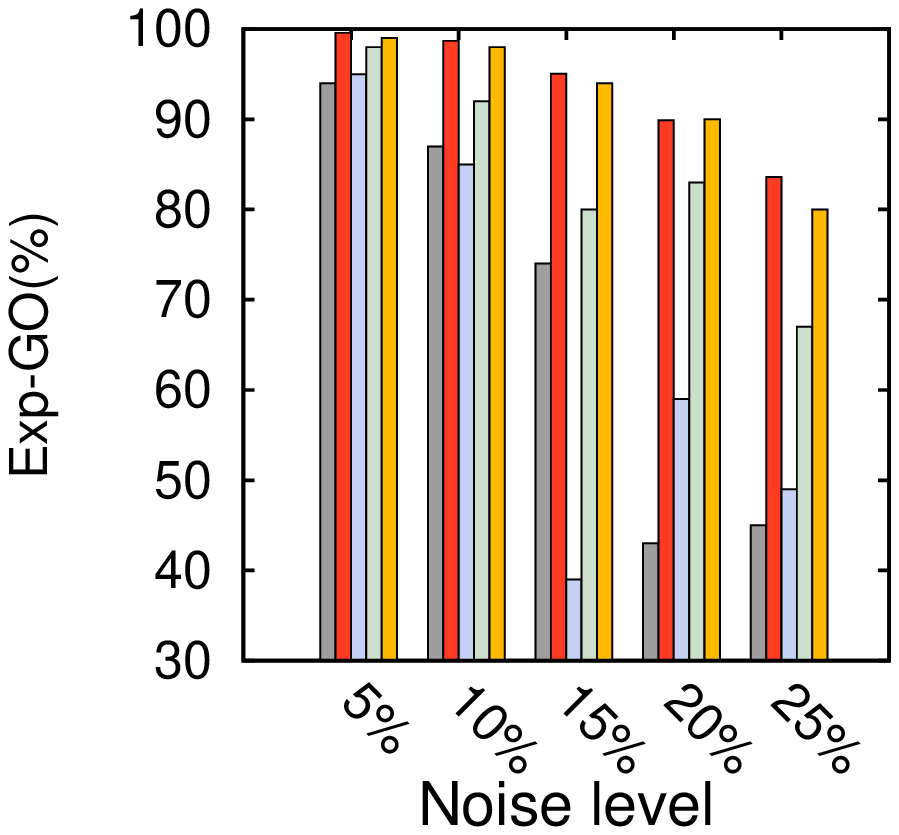}&
\hspace{-0.8cm} \includegraphics[width=0.07\columnwidth]{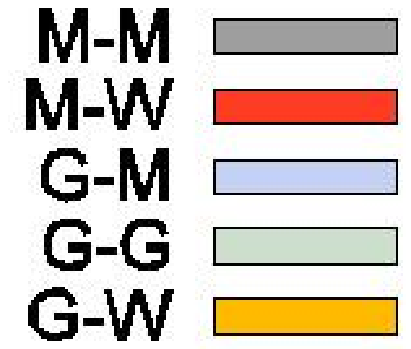}\\
(a)&(b)&(c)&(d) & \\
\end{tabular}
\vspace{-0.2cm}
\caption{Comparison of the five NCF-AS methods on topology-only
alignments of ``synthetic'' (noisy yeast) networks with respect to:
\textbf{(a)} NC, \textbf{(b)} S$^3$, \textbf{(c)} LCCS, and
\textbf{(d)} Exp-GO. }\label{yeast-topo}
\end{figure*}

\noindent\textbf{Topological alignments.} WAVE is always superior to the
existing methods (M-W is superior to M-M, and G-W is superior to G-M
and G-G), for all noise levels and alignment quality measures, under
both MI-GRAAL's and GHOST's NCFs (Figures \ref{overall-ranking} (a)
and \ref{yeast-topo}). 

WAVE in general works better under MI-GRAAL's NCF than under GHOST's
NCF, as M-W is overall superior to G-W. WAVE (at least one of M-W and
G-W) beats both MI-GRAAL and GHOST (all of M-M, G-M, and G-G) in
20/20=100\% of all cases (Figures \ref{overall-ranking} (a) and
\ref{yeast-topo}).

\vspace{0.1cm}

\noindent\textbf{Best alignments.} Under MI-GRAAL's NCF, WAVE is always
superior (M-W is better than M-M), for all noise levels and alignment
quality measures (see the Appendix).

Under GHOST's NCF, WAVE is always superior to MI-GRAAL's AS (G-W is
better than G-M), and WAVE is overall superior to GHOST's AS (G-W is
better than G-G) with respect to two of the four measures (edge-based
S$^3$ and LCCS), while GHOST's AS is superior (G-G is better than G-W)
with respect to the other two measures (node-based NC and Exp-GO)
(see the Appendix). Hence, WAVE and GHOST's AS are comparable overall.

Again, WAVE in general works better under MI-GRAAL's NCF than under
GHOST's, as M-W is overall superior to G-W. WAVE (at least one of M-W
and G-W) beats both MI-GRAAL and GHOST (all of M-M, G-M, and G-G) in
6/10=60\% of cases dealing with the two edge-based measures of
alignment quality, (see the Appendix).

\vspace{-0.5cm}

\subsubsection{Networks With Unknown Node Mapping}

\vspace{-0.2cm}

\begin{figure*}[h]
\centering
\begin{tabular}{ccc}
\includegraphics[width=0.33\columnwidth]{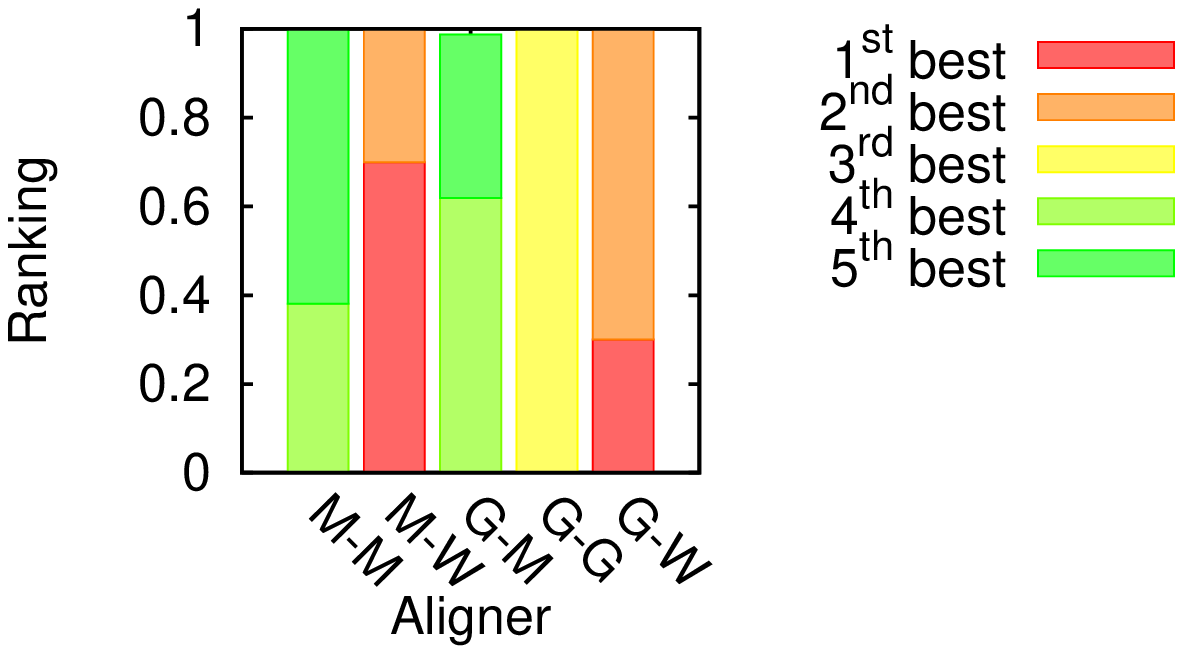} &
\includegraphics[width=0.33\columnwidth]{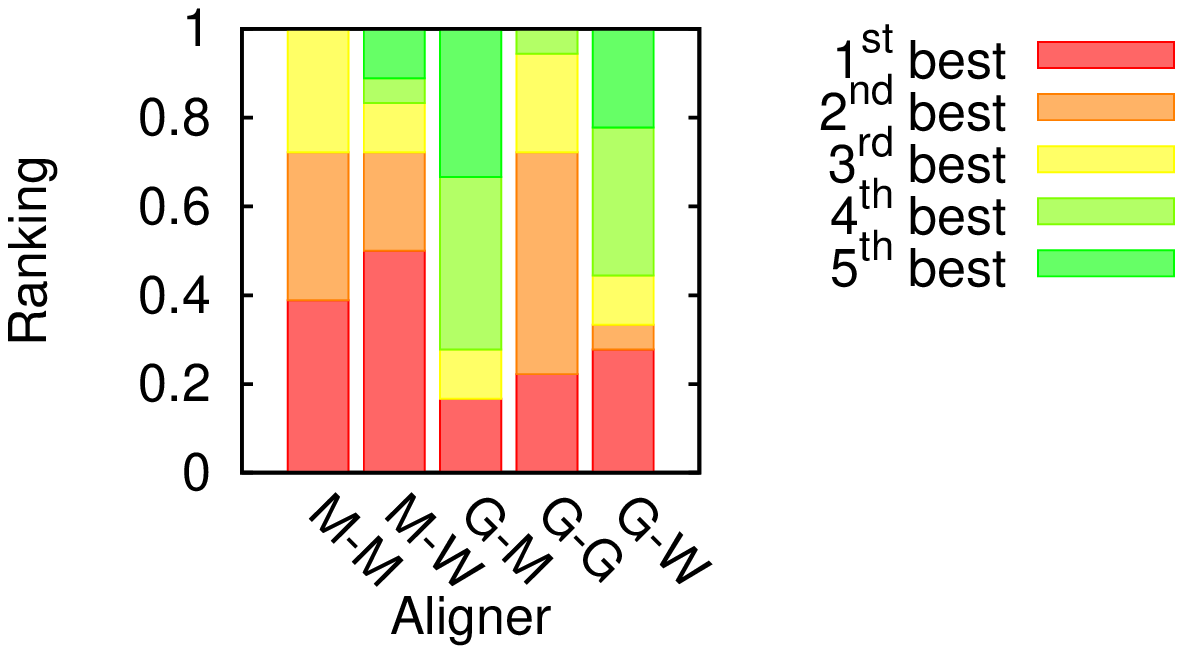} &
\includegraphics[width=0.33\columnwidth]{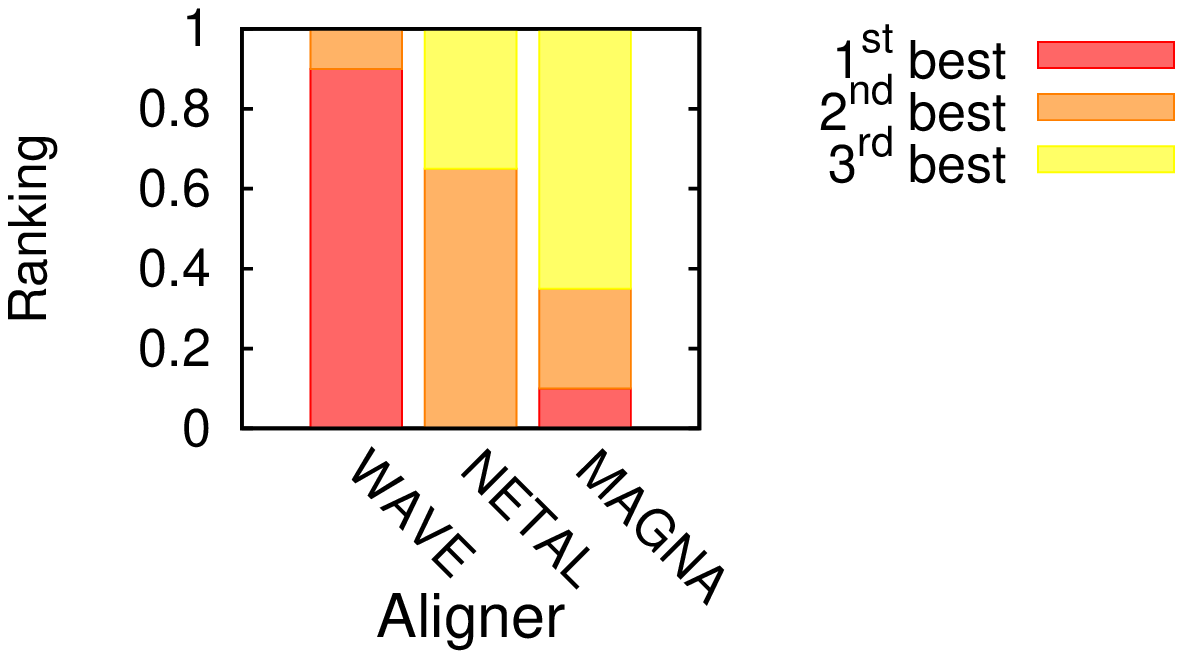} \\
(a)&(b)&(c)\\
\end{tabular}
\vspace{-0.2cm}
\caption{Representative results for overall ranking of each method
over all network pairs in a given data set and over all alignment
quality measures. The ranking is expressed as a percentage of all
cases in which the given method ranks as the $k^{th}$ best method.
\textbf{(a)} Results for the five NCF-AS methods on topology-only
alignments of ``synthetic'' (noisy yeast PPI) networks. For
equivalent results for best alignments, see the Appendix. \textbf{(b)}
Results for the five NCF-AS methods on topology-only alignments of
real-world PPI networks of different species. For equivalent results
for best alignments, see the Appendix. \textbf{(c)} Results for WAVE (the best of M-W and
G-W) against the recent methods (NETAL and MAGNA) on topology-only
alignments of ``synthetic'' (noisy yeast PPI) networks. For
equivalent results for real-world PPI networks, see the Appendix. Details (per
network pair and alignment quality measure) for panels (a)-(c) are
shown in Figures \ref{yeast-topo}-\ref{yeast-topo-mn}, respectively.
Recall that M-M and G-G are MI-GRAAL and
GHOST.}\label{overall-ranking}
\end{figure*}

\begin{figure*}[h]
\centering
\begin{tabular}{cccc}
\hspace{-0.7cm} \includegraphics[width=0.35\columnwidth]{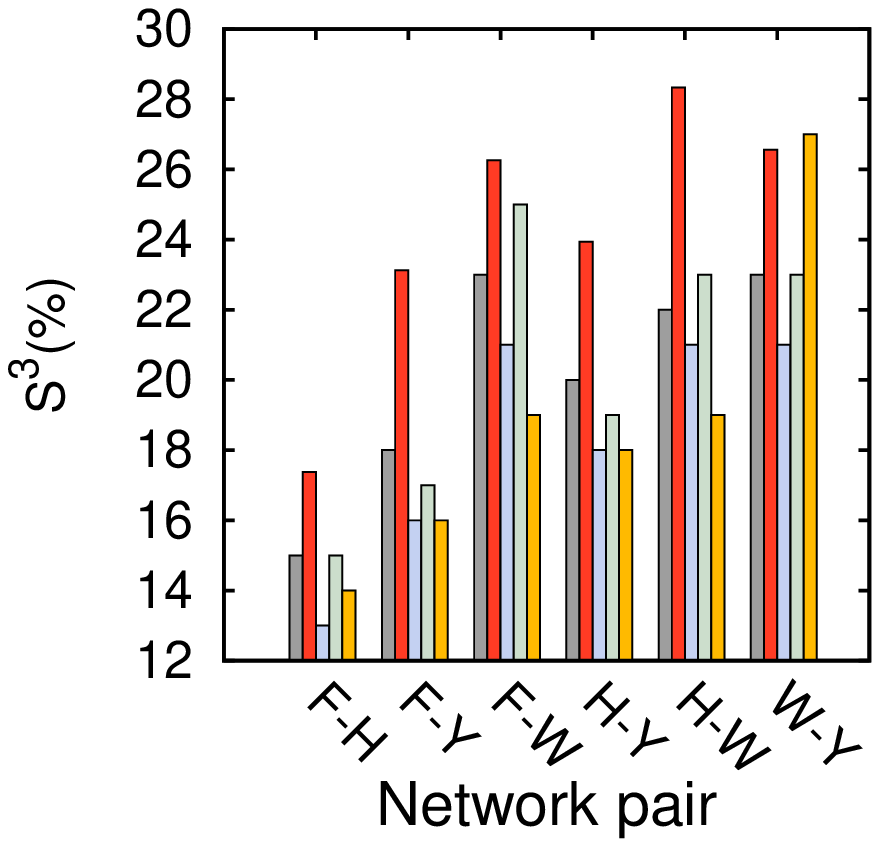}&
\hspace{-1cm} \includegraphics[width=0.35\columnwidth]{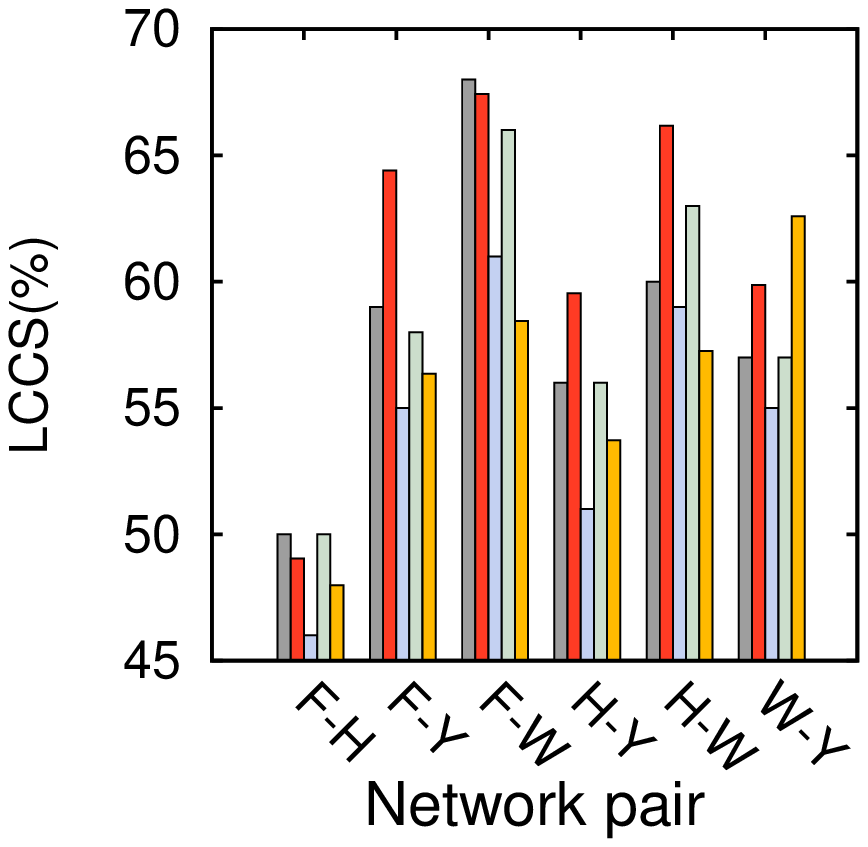}&
\hspace{-1cm} \includegraphics[width=0.35\columnwidth]{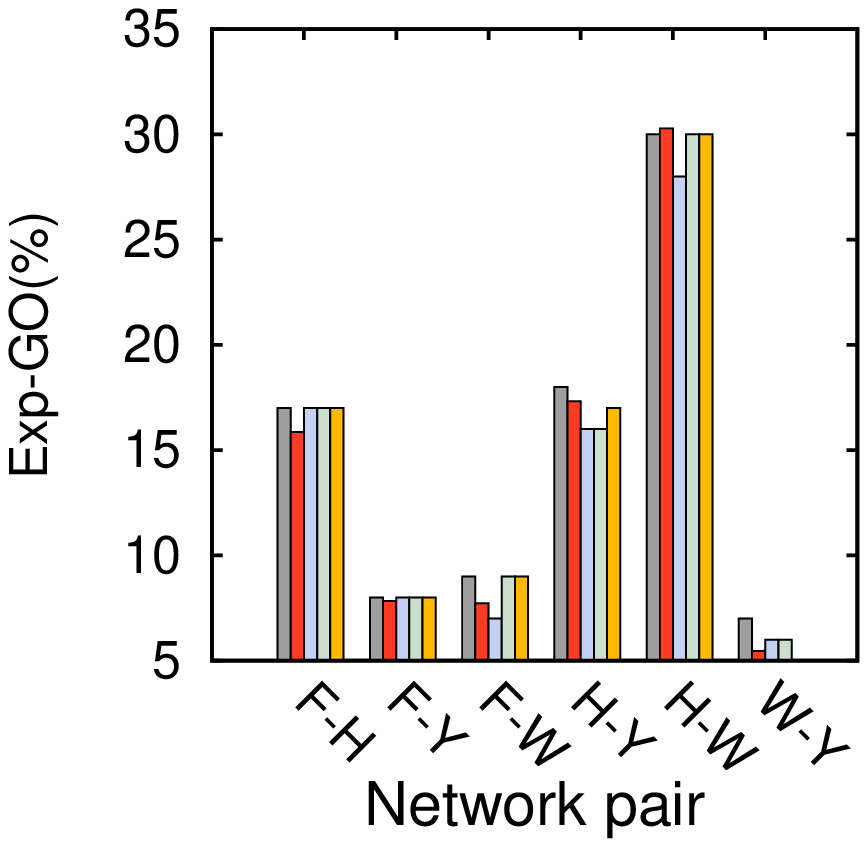}&
\hspace{-0.8cm} \includegraphics[width=0.08\columnwidth]{legend1.eps}\\
(a)&(b)&(c) &\\
\end{tabular}
\vspace{-0.2cm}
\caption{Comparison of the five NCF-AS methods on topology-only
alignments of real-world PPI networks with respect to: \textbf{(a)}
S$^3$, \textbf{(b)} LCCS, and \textbf{(c)} Exp-GO.}\label{real-topo}
\end{figure*}

\noindent\textbf{Topological alignments.} Under MI-GRAAL's NCF, WAVE is always
superior (M-W is better than M-M) with respect to S$^3$, it is almost
always superior with respect to LCCS, and it is sometimes superior
with respect to Exp-GO (Figures \ref{overall-ranking} (b) and
\ref{real-topo}). Hence, under MI-GRAAL's NCF, WAVE seems to be
favored by topological alignment quality measures.

Under GHOST's NCF, WAVE is superior to MI-GRAAL's AS (G-W is better
than G-M) in almost all cases, for each of S$^3$, LCCS, and Exp-GO
(Figures \ref{overall-ranking} (b) and \ref{real-topo}). Also, under
GHOST's NCF, WAVE is overall superior to GHOST's AS (G-W is better
than G-G) with respect to Exp-GO but not with respect to S$^3$ or LCCS
(Figures \ref{overall-ranking} (b) and \ref{real-topo}). 

WAVE in general works better under MI-GRAAL's NCF than under GHOST's
NCF, as M-W is overall superior to G-W. WAVE (at least one of M-W and
G-W) beats both MI-GRAAL and GHOST (all of M-M, G-M, and G-G) in
14/18=78\% of all cases (Figures \ref{overall-ranking} (b) and
\ref{real-topo}).

\vspace{0.1cm}

\noindent\textbf{Best alignments.} Under MI-GRAAL's NCF, WAVE is always
superior (M-W is better than M-M) with respect to S$^3$, and it is
almost always superior with respect to LCCS as well as Exp-GO (see the Appendix). Hence, for best alignments, under MI-GRAAL's NCF, WAVE is
even more superior than for topological alignments only.

Under GHOST's NCF, WAVE is superior to MI-GRAAL's AS (as G-W is better
than G-M) in most cases for each of S$^3$ and Exp-GO, and in some
cases for LCCS. Also, under GHOST's NCF, WAVE is overall superior to
GHOST's AS (G-W is better than G-G) with respect to Exp-GO but not
with respect to S$^3$ or LCCS (see the Appendix).

Again, WAVE works better under MI-GRAAL's NCF than under GHOST's AS,
as M-W is superior to G-W. WAVE (at least one of M-W and G-W) beats
both MI-GRAAL and GHOST (all of M-M, G-M, and G-G) in 13/18=72\% of
all cases (see the Appendix).

The fact that WAVE in general works better under MI-GRAAL's NCF than
under GHOST's NCF further adds to our recent finding that MI-GRAAL's
NCF is superior to other NCFs
\cite{MilenkovicACMBCB2013,Faisal2014a,Crawford2014}.

\vspace{-0.7cm}

\subsection{Comparison of WAVE with Very Recent Methods}

\vspace{-0.3cm}

Here, we compare WAVE (the best of M-W and G-W) with NETAL and MAGNA,
which became available close to completion of our study. As such, we
were unable to include novelties of these methods (and especially
NETAL's NCF) into our methodology. Recall that we compare the three
methods on topology-only alignments, for reasons discussed in Section
\ref{combining}.

\vspace{-0.4cm}

\subsubsection{Networks With Known Node Mapping}

\vspace{-0.2cm}

\begin{figure*}[h]
\centering
\begin{tabular}{ccccc}
\hspace{-0.7cm} \includegraphics[width=0.295\columnwidth]{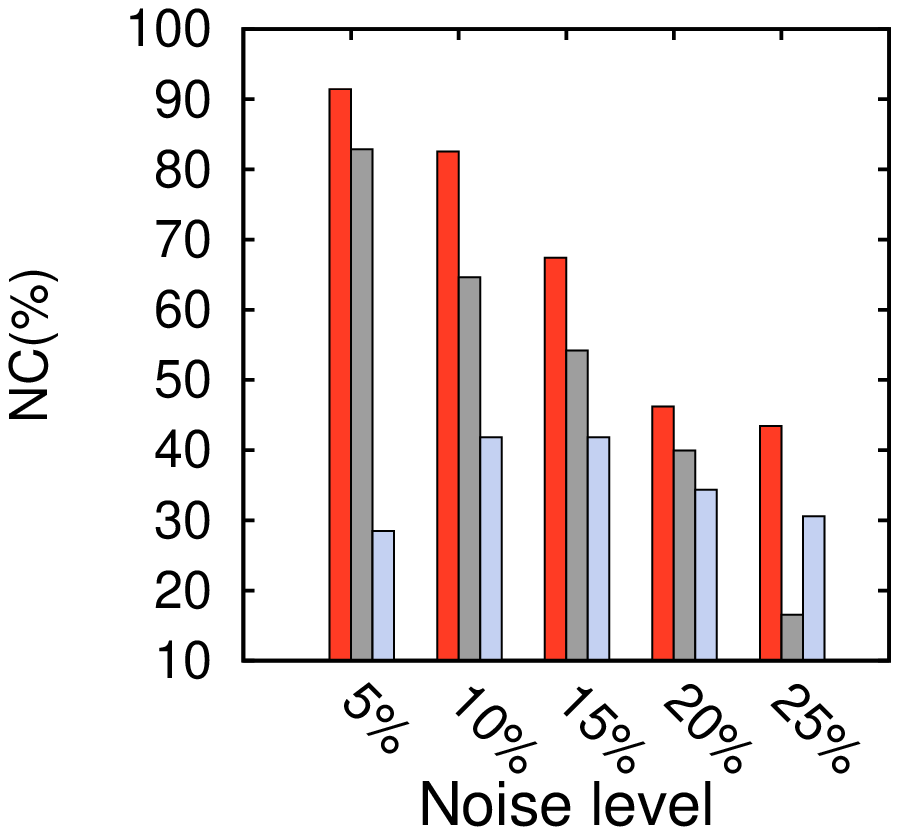}&
\hspace{-1.25cm} \includegraphics[width=0.295\columnwidth]{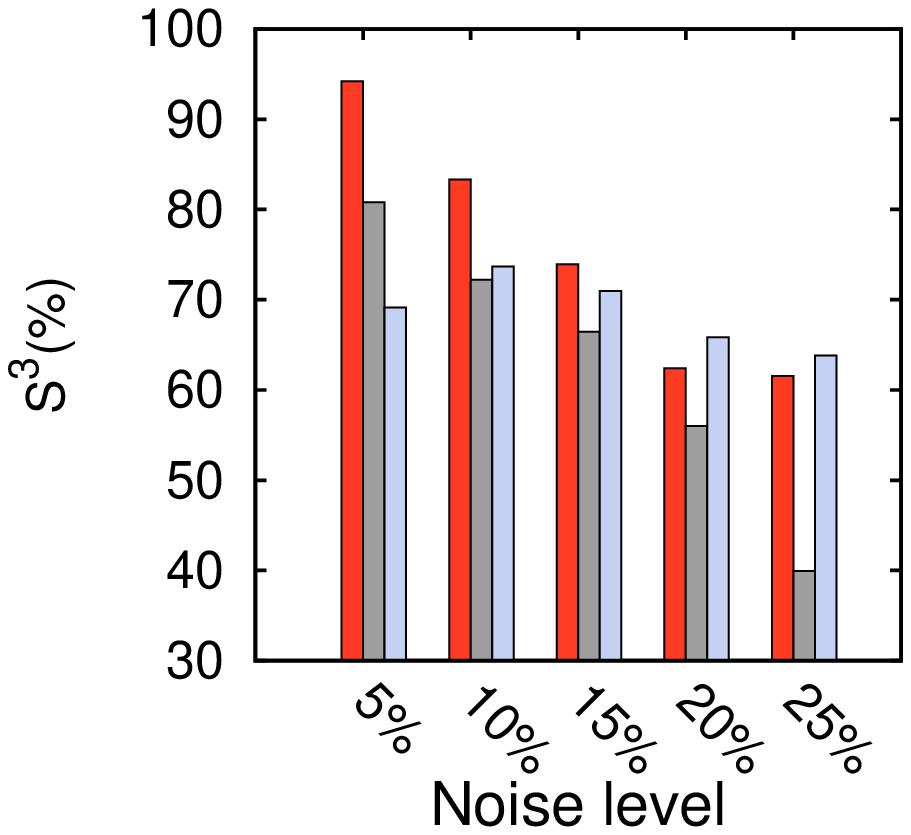}&
\hspace{-1.25cm} \includegraphics[width=0.295\columnwidth]{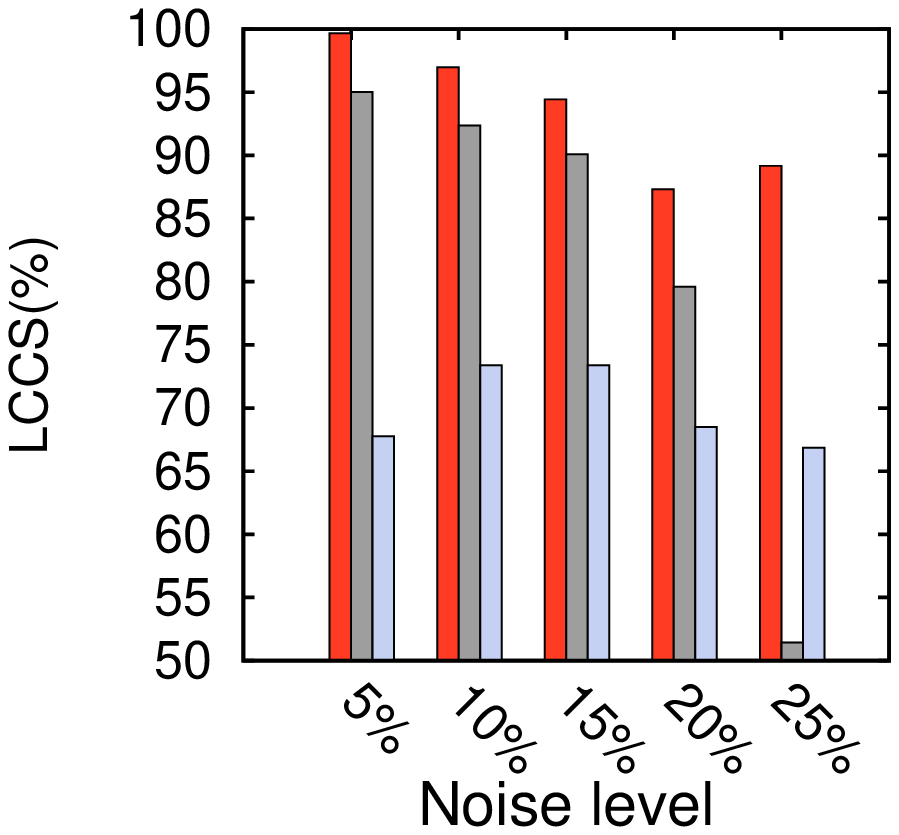}&
\hspace{-1.25cm} \includegraphics[width=0.295\columnwidth]{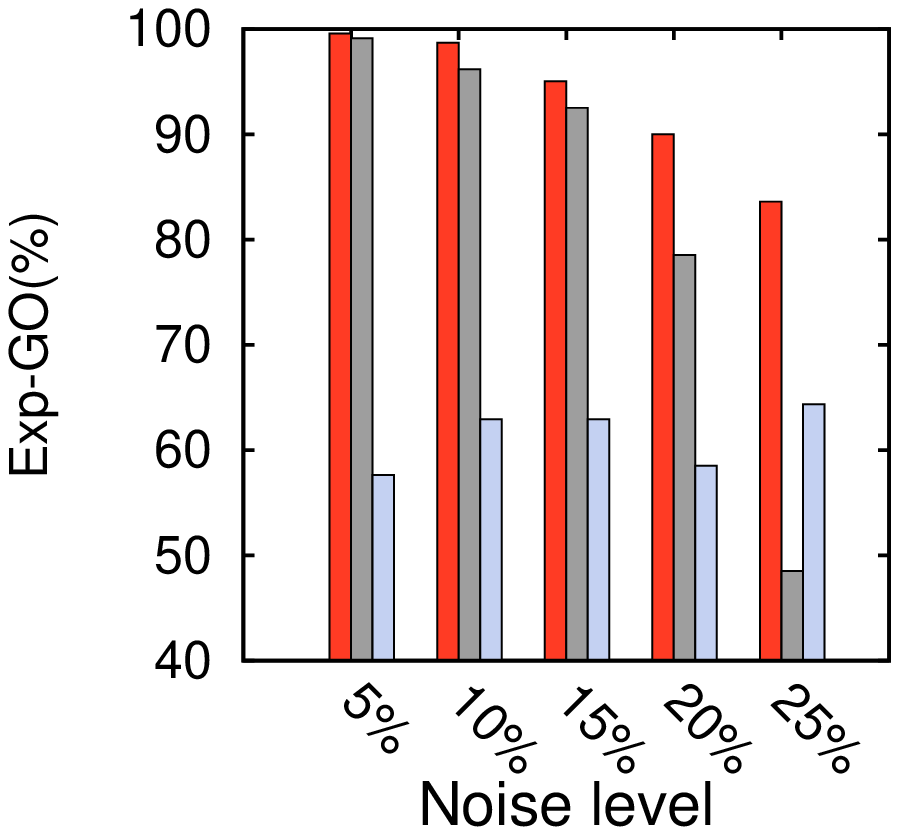}&
\hspace{-0.9cm} \includegraphics[width=0.095\columnwidth]{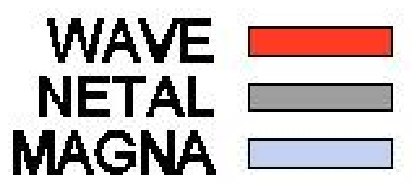}\\
(a)&(b)&(c)&(d)&\\
\end{tabular}
\vspace{-0.2cm}
\caption{Comparison of WAVE (the best of M-W and G-W) with very recent
network alignment methods on topology-only alignments of
``synthetic'' (noisy yeast) networks with respect to: \textbf{(a)}
NC, \textbf{(b)} S$^3$, \textbf{(c)} LCCS, and \textbf{(d)}
Exp-GO.}\label{yeast-topo-mn}
\end{figure*}

WAVE is always superior to both NETAL and MAGNA, for all noise levels
and alignment quality measures (Figures \ref{overall-ranking} (c) and
\ref{yeast-topo-mn}). Only in two out of 20 cases, MAGNA is superior:
with respect to S$^3$ for two largest noise levels. But this is not
surprising, as MAGNA optimizes S$^3$.

\vspace{-0.4cm}

\subsubsection{Networks With Unknown Node Mapping}

\vspace{-0.2cm}

WAVE is always superior to MAGNA, for all noise levels and alignment
quality measures (see the Appendix). Only in one out of 18
cases, MAGNA is superior to WAVE: with respect to S$^3$ for one of the
six network pairs. NETAL is overall superior to the other two
methods, especially with respect to topological alignment quality
measures (S$^3$ and LCCS) (see the Appendix). This could be because NETAL has both
different NCF and AS compared to WAVE, and as such, its superiority
might be a consequence not of its ASs but rather of its NCF. So, if
its NCF was fed into WAVE AS, this could perhaps result in a superior
new method. This possibility of designing a novel superior method
simply by mixing NCF of one method and AS of another method has
already been confirmed on several occasions
\cite{MilenkovicACMBCB2013,Faisal2014a}.

\vspace{-0.75cm}

\vspace{-0.95cm}

\vspace{-1cm}

\vspace{-1cm}

\vspace{0.7cm}

\section{Concluding remarks}
\label{sec:conclusion}

\vspace{-0.3cm}

We have presented WAVE, a general network alignment strategy for
simultaneously optimizing both node conservation and weighted edge
conservation, which can be used with any node cost function or
combination of multiple node cost functions. We have demonstrated
overall superiority of WAVE against existing state-of-the-art
alignment strategies under multiple node cost functions, especially
with respect to topological alignment quality. Moreover, we have
demonstrated that WAVE is comparable or superior even to very recent
approaches that became available only close to completion of our
study, especially on the sythetic network data. This only further
validates the effectiveness of WAVE.

Since WAVE can be combined with any node cost function, doing so for
any recent function might improve its alignment quality. Also, WAVE
itself can be modified to optimize any other measure of node and edge
conservation, which could further improve its accuracy; the measures
that we have used are merely a proof of concept that optimizing both
node and weighted edge conservation can lead to better alignments
compared to optimizing just node conservation (as e.g., MI-GRAAL and
GHOST do) or just unweigted edge conservation (as e.g., MAGNA does).

As more biological network data are becoming available, network
alignment will only continue to gain importance in the computationally
biology domain \cite{Sharan2006,NewSurvey2014,MAGNA}. Further, network
alignment has implications in many domains. For example, it can be
used to de-anonymize online social networks and thus impact privacy
\cite{Narayanan2011}.
Hence, further theoretical improvements that would lead to better
network alignments have a potential to lead to important discoveries
in different fields.

\vspace{-0.5cm}

\section*{Acknowledgements}

\vspace{-0.3cm}

\label{ack}

Funding: the National Science Foundation CCF-1319469 and EAGER
CCF-1243295 grants.

\clearpage

\setcounter{page}{1} \renewcommand{\thepage}{Bibliography Page \arabic{page}}

\small{\bibliography{ref}}

\begin{thebibliography}{57}
\providecommand{\natexlab}[1]{#1}
\providecommand{\url}[1]{\texttt{#1}}
\expandafter\ifx\csname urlstyle\endcsname\relax
  \providecommand{\doi}[1]{doi: #1}\else
  \providecommand{\doi}{doi: \begingroup \urlstyle{rm}\Url}\fi

\bibitem[Altschul et~al.(1990)Altschul, Gish, Miller, Myers, and
  Lipman]{BLASTscore}
Stephen~F Altschul, Warren Gish, Webb Miller, Eugene~W Myers, and David~J
  Lipman.
\newblock Basic local alignment search tool.
\newblock \emph{Journal of Molecular Biology}, 215\penalty0 (3):\penalty0
  403--410, 1990.

\bibitem[Bayati et~al.(2009)Bayati, Gerritsen, Gleich, Saberi, and
  Wang]{symword}
Mohsen Bayati, Margot Gerritsen, David~F Gleich, Amin Saberi, and Ying Wang.
\newblock Algorithms for large, sparse network alignment problems.
\newblock In \emph{Data Mining, 2009. ICDM'09. Ninth IEEE International
  Conference on}, pages 705--710. IEEE, 2009.

\bibitem[Berg and Lassig(2004)]{Berg04}
J.~Berg and M.~Lassig.
\newblock Local graph alignment and motif search in biological networks.
\newblock \emph{Proceedings of the National Academy of Sciences}, 101\penalty0
  (41):\penalty0 14689--14694, 2004.

\bibitem[Berg and Lassig(2006)]{Berg2006}
J.~Berg and M.~Lassig.
\newblock {Cross-species analysis of biological networks by Bayesian
  alignment}.
\newblock \emph{Proceedings of the National Academy of Sciences}, 103\penalty0
  (29):\penalty0 10967--10972, 2006.

\bibitem[Ciriello et~al.(2012)Ciriello, Mina, Guzzi, Cannataro, and
  Guerra]{AlignNemo}
Giovanni Ciriello, Marco Mina, Pietro~H. Guzzi, Mario Cannataro, and Concettina
  Guerra.
\newblock {AlignNemo: A Local Network Alignment Method to Integrate Homology
  and Topology}.
\newblock \emph{{PLOS ONE}}, 7\penalty0 (6), 2012.

\bibitem[Clark and Kalita(2014)]{NewSurvey2014}
Connor Clark and Jugal Kalita.
\newblock A comparison of algorithms for the pairwise alignment of biological
  networks.
\newblock \emph{Bioinformatics}, page btu307, 2014.

\bibitem[Collins et~al.(2007)Collins, Kemmeren, Zhao, Greenblatt, Spencer,
  Holstege, Weissman, and Krogan]{Collins07}
S.R. Collins, P.~Kemmeren, X.C. Zhao, J.F. Greenblatt, F.~Spencer, F.C.P.
  Holstege, J.S. Weissman, and N.J. Krogan.
\newblock Toward a comprehensive atlas of the phyisical interactome of
  {S}accharomyces cerevisiae.
\newblock \emph{Molecular Cell Proteomics}, 6\penalty0 (3):\penalty0 439--450,
  2007.

\bibitem[Conte et~al.(2004)Conte, Foggia, Sansone, and Vento]{pm1}
Donatello Conte, Pasquale Foggia, Carlo Sansone, and Mario Vento.
\newblock Thirty years of graph matching in pattern recognition.
\newblock \emph{International Journal of Pattern Recognition and Artificial
  Intelligence}, 18\penalty0 (03):\penalty0 265--298, 2004.

\bibitem[Cook(1971)]{cook1971}
S.A. Cook.
\newblock The complexity of theorem-proving procedures.
\newblock In \emph{Proceedings of the 3rd Annual ACM Symposium on Theory of
  Computing}, pages 151--158, 1971.

\bibitem[Crawford and Sun(2014)]{Crawford2014}
J.~Crawford and T.~Sun, Y.~Milenkovi{\'c}.
\newblock Fair evaluation of global network aligners.
\newblock 2014.
\newblock arXiv:1407.4824 [q-bio.MN].

\bibitem[De~Magalh{\~a}es et~al.(2009)De~Magalh{\~a}es, Budovsky, Lehmann,
  Costa, Li, Fraifeld, and Church]{GOref}
Jo{\~a}o~Pedro De~Magalh{\~a}es, Arie Budovsky, Gilad Lehmann, Joana Costa,
  Yang Li, Vadim Fraifeld, and George~M Church.
\newblock The human ageing genomic resources: online databases and tools for
  biogerontologists.
\newblock \emph{Aging cell}, 8\penalty0 (1):\penalty0 65--72, 2009.

\bibitem[Duchenne et~al.(2011)Duchenne, Bach, Kweon, and Ponce]{cvcg3}
Olivier Duchenne, Francis Bach, In-So Kweon, and Jean Ponce.
\newblock A tensor-based algorithm for high-order graph matching.
\newblock \emph{Pattern Analysis and Machine Intelligence, IEEE Transactions
  on}, 33\penalty0 (12):\penalty0 2383--2395, 2011.

\bibitem[El-Kebir et~al.(2011)El-Kebir, Heringa, and Klau]{NATALIE2}
Mohammed El-Kebir, Jaap Heringa, and Gunnar~W Klau.
\newblock Lagrangian relaxation applied to sparse global network alignment.
\newblock In \emph{Pattern Recognition in Bioinformatics}, pages 225--236.
  Springer, 2011.

\bibitem[Faisal et~al.(2014)Faisal, Zhao, and Milenkovi{\'c}]{Faisal2014a}
F.~Faisal, H.~Zhao, and T.~Milenkovi{\'c}.
\newblock Global network alignment in the context of aging.
\newblock \emph{{Computational Biology and Bioinformatics, IEEE/ACM
  Transactions on Computational Biology and Bioinformatics}}, PP\penalty0 (99),
  2014.
\newblock ISSN 1545-5963.

\bibitem[Faisal and Milenkovi{\'c}(2014)]{Faisal2014}
Fazle~E Faisal and Tijana Milenkovi{\'c}.
\newblock Dynamic networks reveal key players in aging.
\newblock \emph{Bioinformatics}, 30\penalty0 (12):\penalty0 1721--1729, 2014.

\bibitem[Flannick et~al.(2006)Flannick, Novak, Balaji, Harley, and
  Batzglou]{Flannick2006}
J.~Flannick, A.~Novak, S.S. Balaji, H.M. Harley, and S.~Batzglou.
\newblock Graemlin general and robust alignment of multiple large interaction
  networks.
\newblock \emph{Genome Research}, 16\penalty0 (9):\penalty0 1169--1181, 2006.

\bibitem[Flannick et~al.(2008)Flannick, Novak, Do, Srinivasan, and
  Batzoglou]{Flannick2008}
Jason Flannick, Antal Novak, Chuong~B. Do, Balaji~S. Srinivasan, and Serafim
  Batzoglou.
\newblock {Automatic parameter learning for multiple network alignment}.
\newblock In \emph{Proceedings of the 12th Annual International Conference on
  Research in Computational Molecular Biology}, pages 214--231, 2008.

\bibitem[Guo and Hartemink(2009)]{Guo2009}
X.~Guo and A.J. Hartemink.
\newblock {Domain-oriented edge-based alignment of protein interaction
  networks.}
\newblock \emph{Bioinformatics}, 25\penalty0 (12):\penalty0 i240--1246, 2009.

\bibitem[Hulovatyy et~al.(2014)Hulovatyy, Solava, and
  Milenkovi\'{c}]{Hulovatyy2014}
Y.~Hulovatyy, R.W. Solava, and T.~Milenkovi\'{c}.
\newblock Revealing missing parts of the interactome via link prediction.
\newblock \emph{{PLOS ONE}}, 9\penalty0 (3):\penalty0 e90073, 2014.

\bibitem[Kelley et~al.(2004)Kelley, Yuan, Lewitter, Sharan, Stockwell, and
  Ideker]{PathBlast}
Brian~P. Kelley, Bingbing Yuan, Fran Lewitter, Roded Sharan, Brent~R.
  Stockwell, and Trey Ideker.
\newblock {PathBLAST:} a tool for alignment of protein interaction networks.
\newblock \emph{Nucleic Acids Research}, 32:\penalty0 83--88, 2004.

\bibitem[Klau(2009)]{NATALIE}
G.W. Klau.
\newblock {A new graph-based method for pairwise global network alignment.}
\newblock \emph{BMC Bioinformatics}, 10\penalty0 (Suppl 1):\penalty0 S59, 2009.

\bibitem[Koutra et~al.(2013)Koutra, Tong, and Lubensky]{bilink}
Danai Koutra, Hanghang Tong, and David Lubensky.
\newblock Big-align: Fast bipartite graph alignment.
\newblock In \emph{Data Mining (ICDM), 2013 IEEE 13th International Conference
  on}, pages 389--398. IEEE, 2013.

\bibitem[Koyuturk et~al.(2006)Koyuturk, Kim, Topkara, Subramaniam, Szpankowski,
  and Grama]{Mawish}
M.~Koyuturk, Y.~Kim, U.~Topkara, S.~Subramaniam, W.~Szpankowski, and A.~Grama.
\newblock Pairwise alignment of protein interaction networks.
\newblock \emph{Journal of Computational Biology}, 13\penalty0 (2), 2006.

\bibitem[Kuchaiev and Pr{\v{z}}ulj(2011)]{MIGRAAL}
Oleksii Kuchaiev and Nata{\v{s}}a Pr{\v{z}}ulj.
\newblock Integrative network alignment reveals large regions of global network
  similarity in yeast and human.
\newblock \emph{Bioinformatics}, 27\penalty0 (10):\penalty0 1390--1396, 2011.

\bibitem[Kuchaiev et~al.(2010)Kuchaiev, Milenkovi\'{c}, Memi\v{s}evi\'{c},
  Hayes, and Pr\v{z}ulj]{GRAAL}
Oleksii Kuchaiev, Tijana Milenkovi\'{c}, Vesna Memi\v{s}evi\'{c}, Wayne Hayes,
  and Nata\v{s}a Pr\v{z}ulj.
\newblock {Topological network alignment uncovers biological function and
  phylogeny}.
\newblock \emph{Journal of The Royal Society Interface}, 7\penalty0
  (50):\penalty0 1341--1354, 2010.

\bibitem[Lacoste-Julien et~al.(2013)Lacoste-Julien, Palla, Davies, Kasneci,
  Graepel, and Ghahramani]{sigma}
Simon Lacoste-Julien, Konstantina Palla, Alex Davies, Gjergji Kasneci, Thore
  Graepel, and Zoubin Ghahramani.
\newblock Sigma: Simple greedy matching for aligning large knowledge bases.
\newblock In \emph{Proceedings of the 19th ACM SIGKDD international conference
  on Knowledge discovery and data mining}, pages 572--580. ACM, 2013.

\bibitem[Li et~al.(2009)Li, Tang, Li, and Luo]{rimom}
Juanzi Li, Jie Tang, Yi~Li, and Qiong Luo.
\newblock Rimom: A dynamic multistrategy ontology alignment framework.
\newblock \emph{Knowledge and Data Engineering, IEEE Transactions on},
  21\penalty0 (8):\penalty0 1218--1232, 2009.

\bibitem[Liang et~al.(2006)Liang, Xu, Teng, and Niu]{Liang2006a}
Zhi Liang, Meng Xu, Maikun Teng, and Liwen Niu.
\newblock {NetAlign: A web-based tool for comparison of protein interaction
  networks}.
\newblock \emph{Bioinformatics}, 22\penalty0 (17):\penalty0 2175--2177, 2006.

\bibitem[Liao et~al.(2009)Liao, Lu, Baym, Singh, and Berger]{IsoRankN}
C.~Liao, K.~Lu, M.~Baym, R.~Singh, and B.~Berger.
\newblock Iso{R}ank{N}: {S}pectral methods for global alignment of multiple
  protein networks.
\newblock \emph{Bioinformatics}, 25\penalty0 (12):\penalty0 i253--258, 2009.

\bibitem[Melnik et~al.(2002)Melnik, Garcia-Molina, and Rahm]{flooding}
Sergey Melnik, Hector Garcia-Molina, and Erhard Rahm.
\newblock Similarity flooding: A versatile graph matching algorithm and its
  application to schema matching.
\newblock In \emph{Proc. 18th ICDE Conf.}, 2002.

\bibitem[Memi\v{s}evi\'{c} et~al.(2010)Memi\v{s}evi\'{c}, Milenkovi\'{c}, and
  Pr\v{z}ulj]{Memisevic10b}
V.~Memi\v{s}evi\'{c}, T.~Milenkovi\'{c}, and N.~Pr\v{z}ulj.
\newblock Complementarity of network and sequence information in homologous
  proteins.
\newblock \emph{Journal of Integrative Bioinformatics}, 7\penalty0
  (3):\penalty0 135, 2010.

\bibitem[Milenkovi\'{c} and Pr\v{z}ulj(2008)]{Milenkovic2008}
T.~Milenkovi\'{c} and N.~Pr\v{z}ulj.
\newblock Uncovering biological network function via graphlet degree
  signatures.
\newblock \emph{Cancer Informatics}, 6:\penalty0 257--273, 2008.

\bibitem[Milenkovi\'{c} et~al.(2008)Milenkovi\'{c}, Lai, and
  Pr\v{z}ulj]{GraphCrunch}
T.~Milenkovi\'{c}, J.~Lai, and N.~Pr\v{z}ulj.
\newblock Graph{C}runch: a tool for large network analyses.
\newblock \emph{BMC Bioinformatics}, 9\penalty0 (70), 2008.

\bibitem[Milenkovi\'{c} et~al.(2010{\natexlab{a}})Milenkovi\'{c},
  Memisevi\'{c}, Ganesan, and Pr\v{z}ulj]{MMGP_Roy_Soc_09}
T.~Milenkovi\'{c}, V.~Memisevi\'{c}, A.~K. Ganesan, and N.~Pr\v{z}ulj.
\newblock Systems-level cancer gene identification from protein interaction
  network topology applied to melanogenesis-related interaction networks.
\newblock \emph{Journal of the Royal Society Interface}, 7\penalty0
  (44):\penalty0 423--437, 2010{\natexlab{a}}.

\bibitem[Milenkovi\'{c} et~al.(2010{\natexlab{b}})Milenkovi\'{c}, Ng, Hayes,
  and Pr\v{z}ulj]{HGRAAL}
T.~Milenkovi\'{c}, W.L. Ng, W.~Hayes, and N.~Pr\v{z}ulj.
\newblock Optimal network alignment with graphlet degree vectors.
\newblock \emph{Cancer Informatics}, 9:\penalty0 121--137, 2010{\natexlab{b}}.

\bibitem[Milenkovi\'{c} et~al.(2011)Milenkovi\'{c}, Memi\v{s}evi\'{c}, Bonato,
  and Pr\v{z}ulj]{Milenkovic2011}
T.~Milenkovi\'{c}, V.~Memi\v{s}evi\'{c}, A.~Bonato, and N.~Pr\v{z}ulj.
\newblock Dominating biological networks.
\newblock \emph{{PLOS ONE}}, 6\penalty0 (8):\penalty0 e23016, 2011.

\bibitem[Milenkovi\'{c} et~al.(2013)Milenkovi\'{c}, Zhao, and
  Faisal]{MilenkovicACMBCB2013}
Tijana Milenkovi\'{c}, Han Zhao, and Fazle~E. Faisal.
\newblock Global network alignment in the context of aging.
\newblock In \emph{Proceedings of the International Conference on
  Bioinformatics, Computational Biology and Biomedical Informatics}, BCB'13,
  pages 23:23--23:32. ACM, 2013.

\bibitem[Mina and Guzzi(2014)]{Mina20014}
Marco Mina and Pietro~Hiram Guzzi.
\newblock Improving the robustness of local network alignment: design and
  extensive assessment of a markov clustering-based approach.
\newblock \emph{IEEE/ACM Transactions on Computational Biology and
  Bioinformatics}, 99\penalty0 (PrePrints):\penalty0 1, 2014.

\bibitem[Narayanan et~al.(2011)Narayanan, Shi, and Rubinstein]{Narayanan2011}
A.~Narayanan, E.~Shi, and B.I.P. Rubinstein.
\newblock Link prediction by de-anonymization: How we won the {Kaggle} social
  network challenge.
\newblock In \emph{Proceedings of the 2011 International Joint Conference on
  Neural Networks (IJCNN)}, pages 1825--1834. IEEE, 2011.

\bibitem[Neyshabur et~al.(2013)Neyshabur, Khadem, Hashemifar, and Arab]{NETAL}
Behnam Neyshabur, Ahmadreza Khadem, Somaye Hashemifar, and Seyed~Shahriar Arab.
\newblock {NETAL}: a new graph-based method for global alignment of
  protein-protein interaction networks.
\newblock \emph{Bioinformatics}, 29\penalty0 (13):\penalty0 1654--1662, 2013.

\bibitem[Noma and Cesar(2010)]{cvcg2}
Alexandre Noma and Roberto~Marcondes Cesar.
\newblock Sparse representations for efficient shape matching.
\newblock In \emph{Graphics, Patterns and Images (SIBGRAPI), 2010 23rd SIBGRAPI
  Conference on}, pages 186--192. IEEE, 2010.

\bibitem[Patro and Kingsford(2012)]{GHOST}
R.~Patro and C.~Kingsford.
\newblock {Global network alignment using multiscale spectral signatures}.
\newblock \emph{Bioinformatics}, 28\penalty0 (23):\penalty0 3105--3114, 2012.

\bibitem[Pr\v{z}ulj(2007)]{Przulj06ECCB}
N.~Pr\v{z}ulj.
\newblock Biological network comparison using graphlet degree distribution.
\newblock \emph{Bioinformatics}, 23:\penalty0 e177--e183, 2007.

\bibitem[Saraph and Milenkovi\'{c}(2014)]{MAGNA}
V.~Saraph and T.~Milenkovi\'{c}.
\newblock {MAGNA}: {M}aximizing {A}ccuracy in {G}lobal {N}etwork {A}lignment.
\newblock \emph{Bioinformatics}, page btu409, 2014.

\bibitem[Sharan and Ideker(2006)]{Sharan2006}
R.~Sharan and T.~Ideker.
\newblock Modeling cellular machinery through biological network comparison.
\newblock \emph{Nature Biotechnology}, 24\penalty0 (4):\penalty0 427--433,
  2006.

\bibitem[Sharan et~al.(2005)Sharan, Suthram, Kelley, Kuhn, McCuine, Uetz,
  Sittler, Karp, and Ideker]{Sharan2005}
R.~Sharan, S.~Suthram, R.M. Kelley, T.~Kuhn, S.~McCuine, P.~Uetz, T.~Sittler,
  R.M. Karp, and T.~Ideker.
\newblock {Conserved patterns of protein interaction in multiple species}.
\newblock \emph{Proceedings of the National Academy of Sciences}, 102\penalty0
  (6):\penalty0 1974--1979, 2005.

\bibitem[Singh et~al.(2008)Singh, Xu, and Berger]{Singh2008}
R.~Singh, J.~Xu, and B.~Berger.
\newblock Global alignment of multiple protein interaction networks.
\newblock \emph{In Proceedings of Pacific Symposium on Biocomputing 13}, pages
  303--314, 2008.

\bibitem[Singh et~al.(2007)Singh, Xu, and Berger]{Singh2007}
Rohit Singh, Jinbo Xu, and Bonnie Berger.
\newblock Pairwise global alignment of protein interaction networks by matching
  neighborhood topology.
\newblock In \emph{Research in computational molecular biology}, pages 16--31.
  Springer, 2007.

\bibitem[Smalter et~al.(2008)Smalter, Huan, and Lushington]{chemical}
Aaron Smalter, Jun Huan, and Gerald Lushington.
\newblock Gpm: A graph pattern matching kernel with diffusion for chemical
  compound classification.
\newblock In \emph{BioInformatics and BioEngineering, 2008. BIBE 2008. 8th IEEE
  International Conference on}, pages 1--6. IEEE, 2008.

\bibitem[Solava et~al.(2012)Solava, Michaels, and Milenkovi\'{c}]{Solava2012}
R.W. Solava, R.P. Michaels, and T.~Milenkovi\'{c}.
\newblock Graphlet-based edge clustering reveals pathogen-interacting proteins.
\newblock \emph{Bioinformatics}, 18\penalty0 (28):\penalty0 i480--i486, 2012.

\bibitem[Suchanek et~al.(2011)Suchanek, Abiteboul, and Senellart]{paris}
Fabian~M Suchanek, Serge Abiteboul, and Pierre Senellart.
\newblock Paris: Probabilistic alignment of relations, instances, and schema.
\newblock \emph{Proceedings of the VLDB Endowment}, 5\penalty0 (3):\penalty0
  157--168, 2011.

\bibitem[Torresani et~al.(2008{\natexlab{a}})Torresani, Kolmogorov, and
  Rother]{cvcg1}
Lorenzo Torresani, Vladimir Kolmogorov, and Carsten Rother.
\newblock Feature correspondence via graph matching: Models and global
  optimization.
\newblock In \emph{Computer Vision--ECCV 2008}, pages 596--609. Springer,
  2008{\natexlab{a}}.

\bibitem[Torresani et~al.(2008{\natexlab{b}})Torresani, Kolmogorov, and
  Rother]{cvcg4}
Lorenzo Torresani, Vladimir Kolmogorov, and Carsten Rother.
\newblock Feature correspondence via graph matching: Models and global
  optimization.
\newblock In \emph{Computer Vision--ECCV 2008}, pages 596--609. Springer,
  2008{\natexlab{b}}.

\bibitem[Venkatesan et~al.(2009)Venkatesan, Rual, Vazquez, Stelzl, Lemmens,
  Hirozane-Kishikawa, and \emph{et al.}]{Venkatesan2009}
K.~Venkatesan, J.F. Rual, A.~Vazquez, U.~Stelzl, I.~Lemmens,
  T.~Hirozane-Kishikawa, and \emph{et al.}
\newblock {An empirical framework for binary interactome mapping.}
\newblock \emph{Nature Methods}, 6\penalty0 (1):\penalty0 83--90, 2009.

\bibitem[Zaslavskiy et~al.(2009{\natexlab{a}})Zaslavskiy, Bach, and
  Vert]{GraphM}
M.~Zaslavskiy, F.~Bach, and J.~P. Vert.
\newblock Global alignment of protein-protein interaction networks by graph
  matching methods.
\newblock \emph{Bioinformatics}, 25\penalty0 (12):\penalty0 i259--i267,
  2009{\natexlab{a}}.

\bibitem[Zaslavskiy et~al.(2009{\natexlab{b}})Zaslavskiy, Bach, and Vert]{pm2}
Mikhail Zaslavskiy, Francis Bach, and J-P Vert.
\newblock A path following algorithm for the graph matching problem.
\newblock \emph{Pattern Analysis and Machine Intelligence, IEEE Transactions
  on}, 31\penalty0 (12):\penalty0 2227--2242, 2009{\natexlab{b}}.

\bibitem[Zhang and Tang(2013)]{crosslinking}
Yutao Zhang and Jie Tang.
\newblock Social network integration: Towards constructing the social graph.
\newblock \emph{arXiv preprint arXiv:1311.2670 [cs.SI]}, 2013.

\end{thebibliography}

\clearpage

\end{document}